\documentclass{article}

\usepackage[preprint, nonatbib]{neurips_2020}

\usepackage[utf8]{inputenc} 
\usepackage[T1]{fontenc}    
\usepackage{hyperref}       
\usepackage{url}            
\usepackage{booktabs}       
\usepackage{amsfonts}       
\usepackage{nicefrac}       
\usepackage{microtype}      
\usepackage{graphicx}       
\usepackage{subcaption}

\title{COVIDNet-CT: A Tailored Deep Convolutional Neural Network Design for Detection of COVID-19 Cases from Chest CT Images}

\author{
    Hayden Gunraj\\
    Department of Mechanical and Mechatronics Engineering\\
    University of Waterloo, Canada\\
    \texttt{hayden.gunraj@uwaterloo.ca}\\
    \And
    Linda Wang\\
    Vision and Image Processing Research Group\\
    University of Waterloo, Canada\\
    DarwinAI Corp., Canada\\
    \texttt{linda.wang@uwaterloo.ca}\\
    \And
    Alexander Wong\\
    Vision and Image Processing Research Group\\
    University of Waterloo, Canada\\
    DarwinAI Corp., Canada\\
    \texttt{a28wong@uwaterloo.ca}\\
}

\begin{document}

\maketitle

\begin{abstract}

    The coronavirus disease 2019 (COVID-19) pandemic continues to have a tremendous impact on patients and healthcare systems around the world. In the fight against this novel disease, there is a pressing need for rapid and effective screening tools to identify patients infected with COVID-19, and to this end CT imaging has been proposed as one of the key screening methods which may be used as a complement to RT-PCR testing, particularly in situations where patients undergo routine CT scans for non-COVID-19 related reasons, patients with worsening respiratory status or developing complications that require expedited care, and patients suspected to be COVID-19-positive but have negative RT-PCR test results. Early studies on CT-based screening have reported abnormalities in chest CT images which are characteristic of COVID-19 infection, but these abnormalities may be difficult to distinguish from abnormalities caused by other lung conditions. Motivated by this, in this study we introduce COVIDNet-CT, a deep convolutional neural network architecture that is tailored for detection of COVID-19 cases from chest CT images via a machine-driven design exploration approach. Additionally, we introduce COVIDx-CT, a benchmark CT image dataset derived from CT imaging data collected by the China National Center for Bioinformation comprising 104,009 images across 1,489 patient cases. Furthermore, in the interest of reliability and transparency, we leverage an explainability-driven performance validation strategy to investigate the decision-making behaviour of COVIDNet-CT, and in doing so ensure that COVIDNet-CT makes predictions based on relevant indicators in CT images. Both COVIDNet-CT and the COVIDx-CT dataset are available to the general public in an open-source and open access manner as part of the COVID-Net initiative. While COVIDNet-CT is not yet a production-ready screening solution, we hope that releasing the model and dataset will encourage researchers, clinicians, and citizen data scientists alike to leverage and build upon them.

\end{abstract}

\section{Introduction}

    Coronavirus disease 2019 (COVID-19), caused by severe acute respiratory syndrome coronavirus 2 (SARS-CoV-2), continues to have a tremendous impact on patients and healthcare systems around the world. In the fight against this novel disease, there is a pressing need for fast and effective screening tools to identify patients infected with COVID-19 in order to ensure timely isolation and treatment. Currently, real-time reverse transcription polymerase chain reaction (RT-PCR) testing is the primary means of screening for COVID-19, as it can detect SARS-CoV-2 ribonucleic acid (RNA) in sputum samples collected from the upper respiratory tract~\cite{Wang2020_RTPCR}. While RT-PCR testing for COVID-19 is highly specific, its sensitivity is variable depending on sampling method and time since onset of symptoms~\cite{Yang2020, Li2020_RTPCR, Ai2020}, and some studies have reported relatively low COVID-19 sensitivity~\cite{Fang2020, Li2020_RTPCR}. Moreover, RT-PCR testing is a time-consuming process which is in high demand, leading to possible delays in obtaining test results.

\begin{figure}[ht]
        \centering
        \begin{subfigure}[b]{0.45\textwidth}
            \centering
            \includegraphics[width=\textwidth]{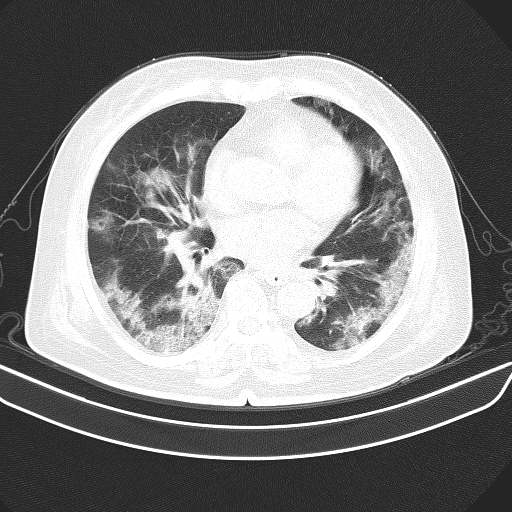}
            \caption{}
        \end{subfigure}
        \hspace{0.05\textwidth}
        \begin{subfigure}[b]{0.45\textwidth}
            \centering
            \includegraphics[width=\textwidth]{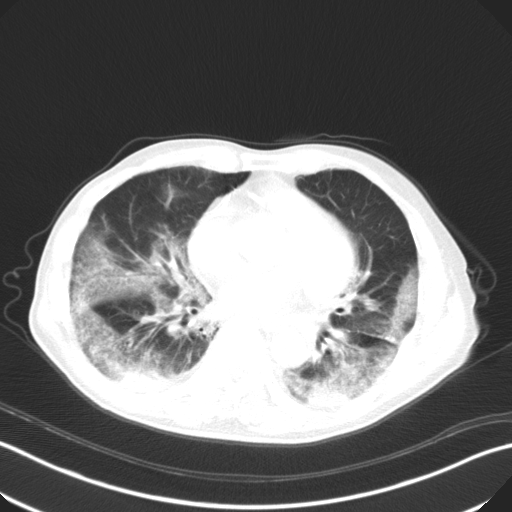}
            \caption{}
        \end{subfigure}
        \caption{Example chest CT abnormalities in (a) a patient with COVID-19 pneumonia, and (b) a patient with non-COVID-19 pneumonia. It can be observed that visual difference in abnormalities between COVID-19 pneumonia and non-COVID-19 penumonia can be quite subtle.}
        \label{fig:comparison}
    \end{figure}

    Chest computed tomography (CT) imaging has been proposed as an alternative screening tool for COVID-19 infection due to its high sensitivity, and may be particularly effective when used as a complement to RT-PCR testing~\cite{Fang2020, Ai2020, Xie2020}. CT imaging saw extensive use during the early stages of the COVID-19 pandemic, particularly in Asia. While cost and resource constraints limit routine CT screening specifically for COVID-19 detection~\cite{ACR}, CT imaging can be especially useful as a screening tool for COVID-19 infection in situations where:
    \begin{itemize}
        \item patients are undergoing routine CT examinations for non-COVID-19 related reasons. For example, CT examinations may be conducted for routine cancer screening, monitoring for elective surgical procedures~\cite{Tian}, and neurological examinations~\cite{Shatri}. Since such CT examinations are being conducted as a routine procedure regardless of COVID-19, there is no additional cost or resource constraints associated with leveraging such examinations for COVID-19 screening as well.
        \item patients have worsening respiratory status or developing complications that require expedited care~\cite{SinghE455}. In such scenarios, immediate treatment of patients may be necessary and thus CT imaging is conducted on the patient for COVID-19 infection while waiting for RT-PCR testing to confirm COVID-19 infection.
        \item patients are suspected to be COVID-19-positive but their RT-PCR tests are negative. For example, patients who have had close contact with confirmed COVID-19 cases and are exhibiting symptoms of the disease are highly suspect, but may have negative RT-PCR results. In these cases, CT imaging may be used to confirm COVID-19 infection pending positive RT-PCR results.
    \end{itemize}
    In early studies, it was found that certain abnormalities in chest CT images are indicative of COVID-19 infection, with ground-glass opacities, patchy shadows, crazy-paving pattern, and consolidation being some of the most commonly reported abnormalities, typically with bilateral involvement~\cite{Guan2020, Wang2020, Chung2020, Pan2020, Fang2020, Ai2020, Xie2020}. Moreover, some studies have found that abnormalities in a patient's chest CT scan due to COVID-19 infection may be present despite a negative RT-PCR test~\cite{Fang2020, Ai2020, Xie2020}. However, as illustrated in Figure~\ref{fig:comparison}, these imaging abnormalities may not be specific to COVID-19 infection, and the visual differences between COVID-19-related abnormalities and other abnormalities can be quite subtle. As a result, the performance of radiologists in distinguishing COVID-19-related abnormalities from abnormalities of other etiology may vary considerably~\cite{Bai2020_Perf, Mei2020}. For radiologists, visual analysis of CT scans is also a time-consuming manual task, particularly when patient volume is high or in large studies.

    In this study, we introduce COVIDNet-CT, a deep convolutional neural network architecture tailored specifically for detection of COVID-19 cases from chest CT images via a machine-driven design exploration approach. We also introduce COVIDx-CT, a benchmark CT image dataset derived from CT imaging data collected by the China National Center for Bioinformation (CNCB)~\cite{cncb} comprising 104,009 images across 1,489 patient cases. Additionally, to investigate the decision-making behaviour of COVIDNet-CT, we perform an explainability-driven performance validation and analysis of its predictions, allowing us to explore the critical visual factors associated with COVID-19 infection while also auditing COVIDNet-CT to ensure that its decisions are based on relevant CT image features. In an effort to encourage continued research and development, COVIDNet-CT and the COVIDx-CT dataset are available to the general public in an open-source and open access manner as part of the COVID-Net~\cite{covidnet,alex2020covidnets} initiative, a global open initiative for accelerating collaborative advancement of artificial intelligence for assisting in the fight against the COVID-19 pandemic.

    The paper is organized as follows. We first discuss related work on deep learning systems for CT-based COVID-19 detection in Section~2. Next, we discuss the construction of the COVIDx-CT dataset, the design strategy used to build COVIDNet-CT, the architecture design of COVIDNet-CT, and the explainability-driven performance validation strategy leveraged to audit COVIDNet-CT in Section~3. Following this, in Section~4, we present and discuss the results of our experiments to evaluate the efficacy and decision-making behaviour of the proposed COVIDNet-CT, as well as a comparison of COVIDNet-CT to existing deep neural network architectures for the task of COVID-19 detection on chest CT images. Finally, we draw conclusions and discuss future directions in Section~5.

\section{Related work}

    A number of studies have proposed deep learning systems based on chest CT imaging to distinguish COVID-19 cases from non-COVID-19 cases (which may include both normal and abnormal cases)~\cite{Mei2020, Xu2020, Bai2020_Aug, Li2020, Ardakani2020, Shah2020, Chen2020, Zheng2020, Jin2020, cncb, Jin2020_2, Song2020, Wang2020_Screen}. Many of the proposed systems further identify non-COVID-19 cases as normal~\cite{Xu2020, cncb, Jin2020_2, Song2020}, non-COVID-19 pneumonia (e.g., bacterial pneumonia, viral pneumonia, community-acquired pneumonia (CAP), etc.)~\cite{Bai2020_Aug, Li2020, Ardakani2020, Xu2020, cncb, Song2020, Wang2020_Screen}, or non-pneumonia~\cite{Li2020}. Additionally, some of the proposed systems require lung and/or lung lesion segmentation~\cite{Bai2020_Aug, Mei2020, Chen2020, Zheng2020, cncb}, which necessitates either a segmentation stage in the proposed systems or manual segmentation by radiologists. To the best of the authors' knowledge, the proposed COVIDNet-CT deep neural network architecture is the first to be built using a machine-driven design exploration strategy specifically for COVID-19 detection from chest CT images.

    Explainability methods have been leveraged in some studies to investigate the relationship between imaging features and network predictions. Bai et al.~\cite{Bai2020_Aug} and Jin et al.~\cite{Jin2020_2} visualized importance variations in chest CT images using Gradient-weighted Class Activation Mapping (Grad-CAM)~\cite{gradcam}. Similarly, Mei et al.~\cite{Mei2020} created heatmaps of COVID-19 infection probabilities within receptive fields by upsampling their network's predictions to match chest CT image dimensions. Zhang et al.~\cite{cncb} examined the correlation between key clinical parameters and segmented lung lesion features in chest CT images. To the best of the authors' knowledge, this is the first study to perform explainability-driven performance validation on deep neural networks for COVID-19 detection from chest CT images using an explainability method geared towards identifying specific critical factors as opposed to more general heatmaps that illustrate importance variations within an image.

\section{Methods}

    In this section, we will first describe the methodology behind the construction of the COVIDx-CT open access benchmark dataset with which we train and evaluate the proposed COVIDNet-CT deep neural network architecture.  Next, we describe the methodology behind the creation of the proposed COVIDNet-CT deep neural network architecture via a machine-driven design exploration strategy, as well as the resulting deep neural network architecture. Furthermore, we discuss the implementation details as well as training strategy used to train COVIDNet-CT. Finally, we describe in detail the strategy for explainability-driven performance validation of COVIDNet-CT.

\subsection{COVIDx-CT dataset}

    \begin{figure}[ht]
        \centering
        \begin{subfigure}[b]{1\textwidth}
            \centering
            \includegraphics[width=\textwidth]{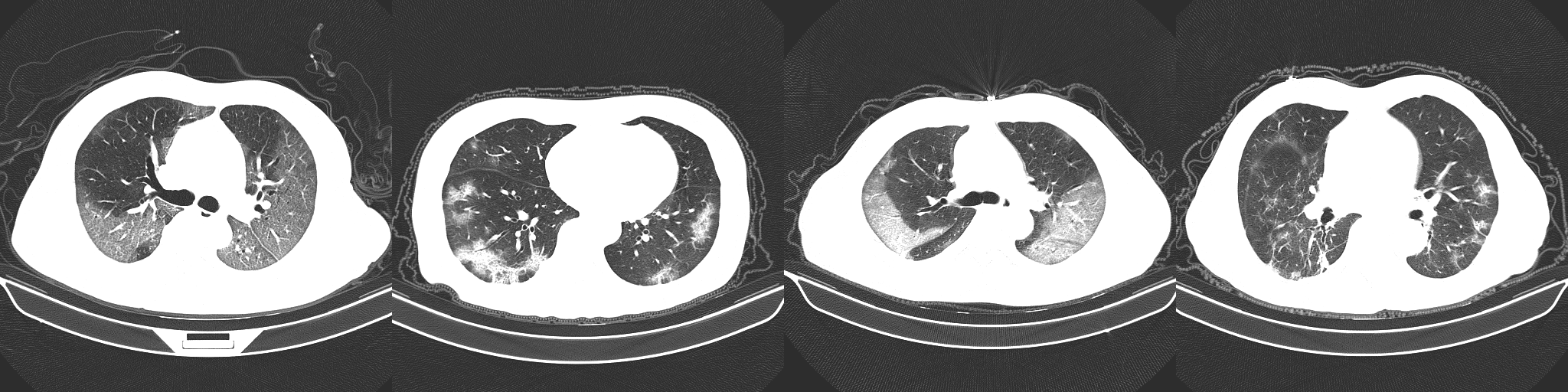}
            \caption{}
            \medskip
        \end{subfigure}
        \begin{subfigure}[b]{1\textwidth}
            \centering
            \includegraphics[width=\textwidth]{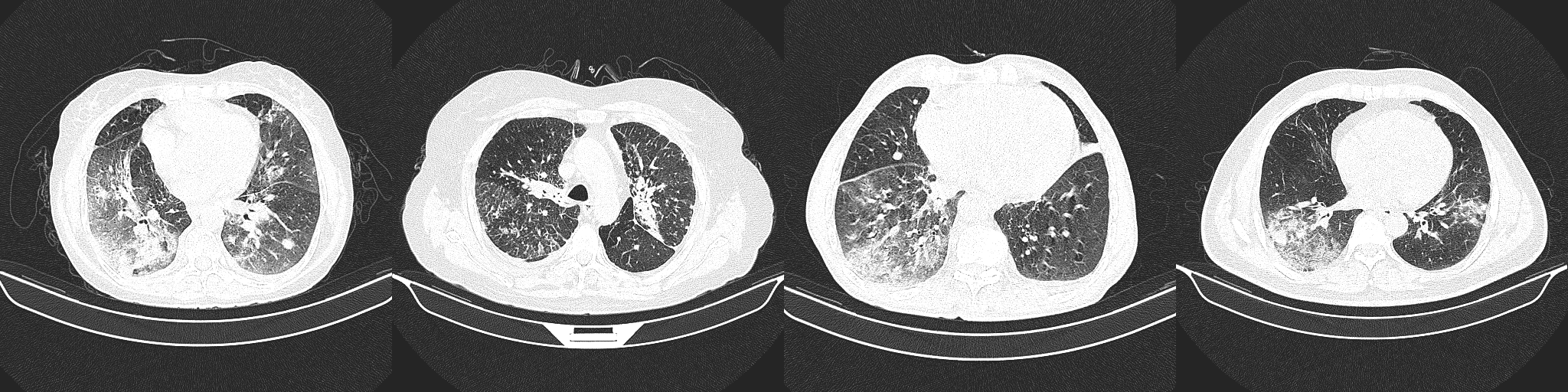}
            \caption{}
            \medskip
        \end{subfigure}
        \begin{subfigure}[b]{1\textwidth}
            \centering
            \includegraphics[width=\textwidth]{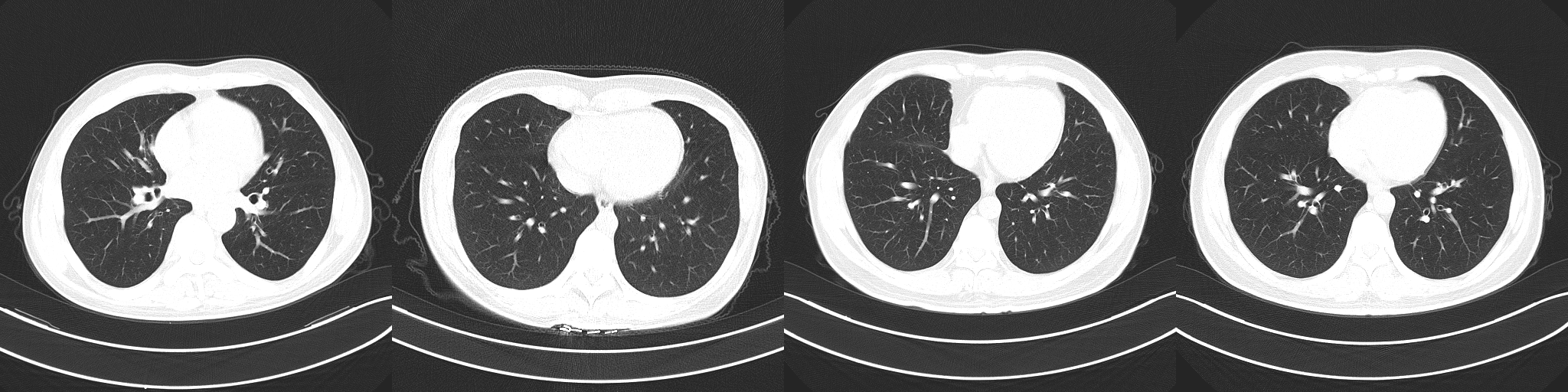}
            \caption{}
        \end{subfigure}
        \caption{Example chest CT images from the COVIDx-CT dataset, illustrating (a) COVID-19 pneumonia cases, (b) non-COVID-19 pneumonia cases, and (c) normal control cases.}
        \label{fig:inf_types}
    \end{figure}

\begin{figure}[ht]
        \centering
        \begin{subfigure}[b]{0.465\textwidth}
            \centering
            \includegraphics[width=\textwidth]{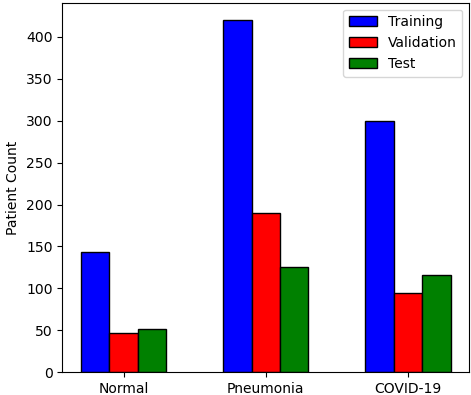}
            \caption{}
        \end{subfigure}
        \hfill
        \begin{subfigure}[b]{0.48\textwidth}
            \centering
            \includegraphics[width=\textwidth]{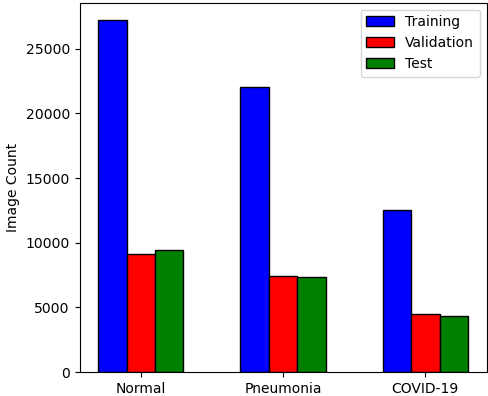}
            \caption{}
        \end{subfigure}
        \caption{Distribution of the COVIDx-CT dataset amongst training, validation, and test sets by (a) patient count and (b) image count.}
        \label{fig:distribution}
    \end{figure}

    To build the proposed COVIDNet-CT, we constructed a dataset of 104,009 chest CT images across 1,489 patient cases, which we refer to as COVIDx-CT. To generate the COVIDx-CT dataset, we leverage the CT imaging data collected by the CNCB~\cite{cncb}, which is comprised of chest CT examinations from different hospital cohorts across China as part of the China Consortium of Chest CT Image Investigation (CC-CCII). More specifically, the CT imaging data consists of chest CT volumes across three different infection types: novel coronavirus pneumonia due to SARS-CoV-2 viral infection (NCP), common pneumonia (CP), and normal controls. Figure~\ref{fig:inf_types} shows example CT images for each of the infection types from the constructed COVIDx-CT dataset. For NCP and CP CT volumes, slices marked as containing lung abnormalities were leveraged. Additionally, we excluded CT volumes where the background had been removed to leave segmented lung regions, as the contrast between the background and segmented lung regions can lead to model biases. Finally, we split the COVIDx-CT dataset into training, validation, and test sets, using an approximate 60\%-20\%-20\% split for training, validation, and test, respectively. These sets were constructed such that each patient belongs to a single set. Figure~\ref{fig:distribution} shows the distribution of patient cases and images in the COVIDx-CT dataset amongst the different infection types and dataset splits.

\subsection{Machine-driven design exploration}

    Inspired by~\cite{covidnet}, a machine-driven design exploration strategy was leveraged to create the proposed COVIDNet-CT. More specifically, machine-driven design exploration involves the automatic exploration of possible network architecture designs and identifies the optimal microarchitecture and macroarchitecture patterns with which to build the deep neural network architecture. As discussed in~\cite{covidnet}, the use of machine-driven design exploration allows for greater flexibility and granularity in the design process as compared to manual architecture design, and ensures that the resulting architecture satisfies the given operational requirements.  As such, a machine-driven design exploration approach would enable the creation of a tailored deep convolutional neural network catered specifically for the purpose of COVID-19 detection from chest CT images in a way that satisfies sensitivity and positive predictive value (PPV) requirements while minimizing computational and architectural complexity to enable widespread adoption in clinical environments where computing resources may be limited.

    More specifically, in this study we leverage the concept of generative synthesis~\cite{gensynth} as our machine-driven design exploration strategy, where the problem of identifying a tailored deep neural network architecture for the task and data at hand is formulated as a constrained optimization problem based on a universal performance function $\mathcal{U}$ (e.g.,~\cite{wong2019netscore}) and a set of quantitative constraints based on operational requirements related to the task and data at hand. This constrained optimization problem is then solved via an iterative strategy, initialized with the data at hand, an initial network design prototype, and the set of quantitative constraints. Here, we specify two key operational requirements as quantitative constraints during the machine-driven design exploration process: (i) COVID-19 sensitivity $\geq$ 95\% on the COVIDx-CT validation dataset, and (ii) COVID-19 PPV $\geq$ 95\% on the COVIDx-CT validation dataset. These operational requirements were specified in order to ensure low false-negative and false-positive rates respectively. For the initial network design prototype, we leveraged residual architecture design principles~\cite{resnet,resnetv2}, as they have been shown to enable reliable deep architectures which are easier to train to high performance. Furthermore, the output of the initial network design prototype is a softmax layer corresponding to the following prediction categories: (i) no infection (normal), (ii) non-COVID-19 pneumonia, and (iii) COVID-19 viral pneumonia.

\subsection{Network architecture}

\begin{figure}[ht]
        \centering
        \includegraphics[width=\textwidth]{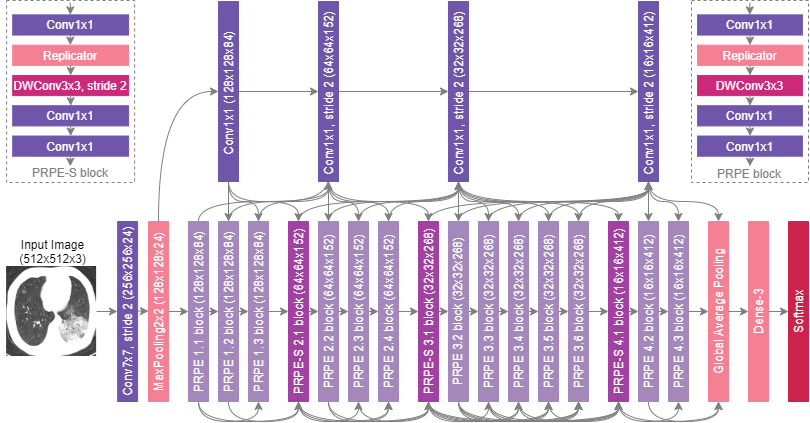}
        \caption{The proposed COVIDNet-CT architecture design via machine-driven design exploration.  Notable characteristics include high architectural diversity, selective long-range connectivity, and lightweight design patterns (e.g., PRPE and PRPE-S patterns).}
        \label{fig:arch}
    \end{figure}

    The proposed COVIDNet-CT architecture is shown in Figure~\ref{fig:arch}, and is publicly available at https://github.com/haydengunraj/COVIDNet-CT. As can be seen, the network architecture produced via a machine-driven design exploration strategy exhibits high architectural diversity as evident by the heterogeneous composition of conventional spatial convolution layers, pointwise convolutional layers, and depthwise convolution layers in a way that strikes a balance between accuracy and architectural and computational complexity. Further evidence of the high architectural diversity of the COVIDNet-CT architecture is the large microarchitecture design variances within each layer of the network (as seen by the tensor configurations of the individual layers shown in Figure~\ref{fig:arch}). Furthermore, the machine-driven design exploration strategy made heavy use of unstrided and strided projection-replication-projection-expansion design patterns (which we denote as PRPE and PRPE-S for unstrided and strided patterns, respectively) consisting of a projection to lower channel dimensionality via pointwise convolutions, a replication of the projections to increase channel dimensionality efficiently, an efficient spatial feature representation via depthwise convolutions (unstrided and strided for PRPE and PRPE-S, respectively), a projection to lower channel dimensionality via pointwise convolutions, and finally an expansion of channel dimensionality conducted by pointwise convolutions. The use of lightweight design patterns such as PRPE and PRPE-S enables COVIDNet-CT to achieve high computational efficiency while maintaining high representational capacity. While these design patterns may be difficult and time-consuming to design manually, machine-driven design allows for these fine-grained design patterns to be rapidly and automatically discovered. Finally, selective long-range connectivity can be observed in the proposed COVIDNet-CT architecture, which enables greater representational capabilities in a more efficient manner than densely-connected deep neural network architectures.

\subsection{Implementation details}
\label{implementation}
\begin{figure}[ht]
        \centering
        \begin{subfigure}[b]{0.45\textwidth}
            \centering
            \includegraphics[width=\textwidth]{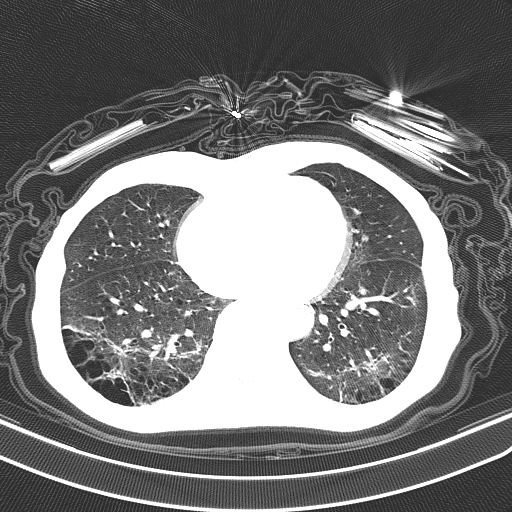}
            \caption{}
        \end{subfigure}
        \hspace{0.05\textwidth}
        \begin{subfigure}[b]{0.45\textwidth}
            \centering
            \includegraphics[width=\textwidth]{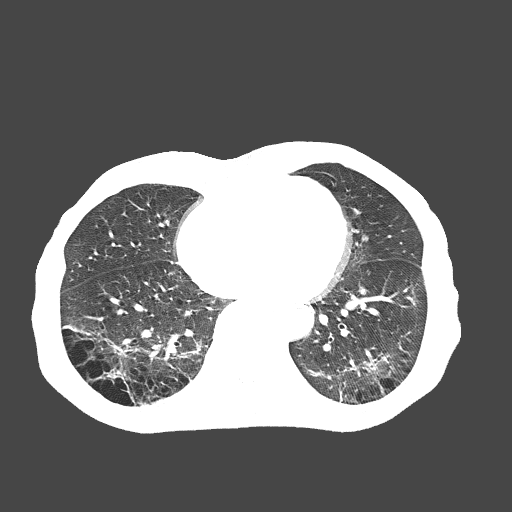}
            \caption{}
        \end{subfigure}
        \caption{Example COVID-19 case before and after removal of irrelevant visual indicators as part of data augmentation. In (a), a number of irrelevant visual indicators are present, such as the patient table of the CT scanner as well as imaging artifacts. After removing these irrelevant indicators, the image in (b) is obtained.}
        \label{fig:exclusion}
    \end{figure}

    The proposed COVIDNet-CT was pre-trained on the ImageNet~\cite{imagenet} dataset and then trained on the COVIDx-CT dataset via stochastic gradient descent with momentum~\cite{momentum}. The hyperparameters used for training are as follows: learning rate=5e-3, momentum=0.9, number of epochs=17, batch size=8. Data augmentation was applied with the following augmentation types: cropping box jitter, rotation, horizontal and vertical shear, horizontal flip, and intensity shift and scaling. In initial experiments, it was found via explainability-driven performance validation (see Section~\ref{xai} for more details on the methodology) that erroneous indicators in the CT images (e.g., patient tables of the CT scanners, imaging artifacts, etc.) were being leveraged by the network to make predictions. To help prevent this behaviour, we introduce an additional augmentation which removes any visual indicators which lie outside of the patient's body, as illustrated in Figure~\ref{fig:exclusion}. Finally, we adopt a batch re-balancing strategy similar to that employed in~\cite{covidnet} to ensure a balanced distribution of each infection type at the batch level. The proposed COVIDNet-CT was implemented, trained, and evaluated using the TensorFlow deep learning library~\cite{tensorflow}.

\subsection{Explainability-driven performance validation of COVIDNet-CT}
\label{xai}

    While scalar performance metrics are a valuable quantitative method for evaluating deep neural networks, they are incapable of explaining a network's decision-making behaviour. In clinical applications, the ability to understand how a deep neural network makes decisions is critical, as these decisions may ultimately affect the health of patients. Motivated by this, we audit COVIDNet-CT via an explainability-driven performance analysis strategy in order to better understand which CT imaging features are critical to its detection decisions. Moreover, by leveraging explainability, we can ensure that COVIDNet-CT is making decisions based on relevant information in CT images rather than erroneously basing its decisions on irrelevant factors (as we have seen in initial experiments as described in Section~\ref{implementation}). In this study, we leverage GSInquire~\cite{gsinquire} as the explainability method of choice for explainability-driven performance validation to visualize critical factors in CT images. GSInquire leverages the generative synthesis strategy~\cite{gensynth} that was employed for machine-driven design exploration, and was previously shown quantitatively to provide explanations that better reflect the decision-making process of deep neural networks when compared to other state-of-the-art explainability methods~\cite{gsinquire}. Unlike approaches that generate heatmaps pertaining to importance variations within an image, GSInquire can identify specific critical factors within an image that have the greatest impact on the decision-making process.

\section{Experimental results}

    The proposed COVIDNet-CT was analysed both quantitatively and qualitatively to evaluate its detection performance and decision-making behaviour. In particular, we report quantitative performance metrics for the COVIDx-CT test set and qualitatively examine critical factors in the CT images as identified by GSInquire for explainability-driven performance validation.

\subsection{Quantitative results}

    We quantitatively evaluate the performance of the proposed COVIDNet-CT on the COVIDx-CT dataset. For this dataset, we computed the test accuracy as well as sensitivity and PPV for each infection type at the image level. The test accuracy, architectural complexity (in terms of number of parameters), and computational complexity (in terms of number of floating-point operations (FLOPs)) of COVIDNet-CT are shown in Table~\ref{table:comp}. As shown, COVIDNet-CT achieves a relatively high test accuracy of 99.1\% while having relatively low architectural and computational complexity. This highlights one of the benefits of leveraging machine-driven design exploration for identifying the optimal macroarchitecture and microarchitecture designs for building a deep neural network architecture tailored for the task and data at hand. In the case of COVIDNet-CT, the result is a highly accurate yet highly efficient deep neural network architecture that is suitable for scenarios where computational resources are a limiting factor. In clinical scenarios, such architectures may also be suitable for use in embedded devices.

    \begin{table}[ht]
        \caption{Comparison of parameters, FLOPs, and accuracy (image-level) for tested network architectures on the COVIDx-CT dataset. Best results highlighted in \textbf{bold}.}
        \medskip
        \label{table:comp}
        \centering
        \begin{tabular}{lllll}
            \toprule
            Architecture & Parameters (M) & FLOPs (G) & Accuracy (\%) \\
            \midrule
            ResNet-50~\cite{resnetv2} & 23.55 & 42.72 & 98.7 \\
            COVIDNet-CT & \textbf{1.40} & \textbf{4.18} & \textbf{99.1} \\
            \bottomrule
        \end{tabular}
    \end{table}

    \begin{table}[ht]
        \caption{Sensitivity for each infection type at the image level on the COVIDx-CT dataset. Best results highlighted in \textbf{bold}.}
        \medskip
        \label{table:sens}
        \centering
        \begin{tabular}{llll} \hline
            \toprule
            \multicolumn{4}{c}{Sensitivity (\%)} \\
            \midrule
            Architecture & Normal & Non-COVID-19 & COVID-19 \\
            \midrule
            ResNet-50~\cite{resnetv2} & 99.9 & 98.7 & 96.2 \\
            COVIDNet-CT & \textbf{100.0} & \textbf{99.0} & \textbf{97.3} \\
            \bottomrule
        \end{tabular}
    \end{table}

    \begin{table}[ht]
        \caption{Positive predictive value (PPV) for each infection type at the image level on the COVIDx-CT dataset. Best results highlighted in \textbf{bold}.}
        \medskip
        \label{table:ppv}
        \centering
        \begin{tabular}{llll}
            \toprule
            \multicolumn{4}{c}{PPV (\%)} \\
            \midrule
            Architecture & Normal & Non-COVID-19 & COVID-19 \\
            \midrule
            ResNet-50~\cite{resnetv2} & 99.3 & 97.8 & 99.1 \\
            COVIDNet-CT & \textbf{99.4} & \textbf{98.4} & \textbf{99.7} \\
            \bottomrule
        \end{tabular}
    \end{table}

    \begin{figure}[!ht]
        \centering
        \includegraphics[width=0.65\textwidth]{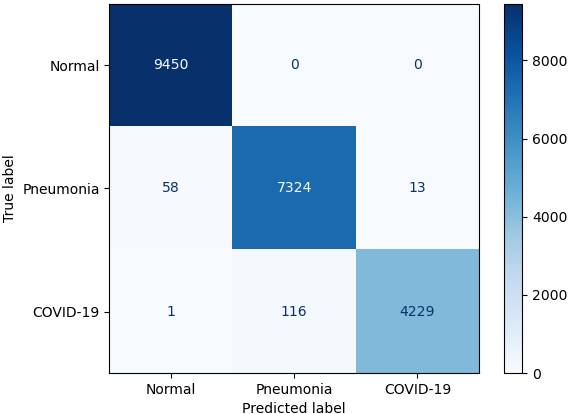}
        \caption{Confusion matrix for COVIDNet-CT on the COVIDx-CT test dataset.}
        \label{fig:confusion}
    \end{figure}

    We next examine the sensitivity and PPV for each infection type in Table~\ref{table:sens} and Table~\ref{table:ppv} respectively, as well as how these statistics could impact the efficacy of COVIDNet-CT in a clinical environment. In Table~\ref{table:sens}, we observe that COVIDNet-CT achieves good COVID-19 sensitivity (97.3\%), which ensures that a low proportion of COVID-19 cases are incorrectly classified as non-COVID-19 pneumonia or normal cases. Moreover, given that RT-PCR testing is highly specific, we want to ensure that COVIDNet-CT has high sensitivity in order to effectively complement RT-PCR testing. Next, in Table~\ref{table:ppv}, we observe that COVIDNet-CT also achieves a high COVID-19 PPV, thereby ensuring a low proportion of false-positive predictions which could cause an unnecessary burden on the healthcare system in the form of isolation, testing, and treatment. Examining Figure~\ref{fig:confusion}, we observe that COVIDNet-CT is extremely effective at distinguishing COVID-19 cases from normal control cases, and is capable of distinguishing non-COVID-19 pneumonia cases from COVID-19 cases for the vast majority of these cases. Interestingly, while some COVID-19 cases are incorrectly classified as non-COVID-19 pneumonia cases, far fewer non-COVID-19 cases are misclassified as COVID-19 cases. Based on these results, it is shown that COVIDNet-CT could be used as an effective standalone screening tool for COVID-19 patients, and could also be used effectively in conjunction with RT-PCR testing. However, we note that COVIDNet-CT is trained on images from a single data collection~\cite{cncb}, and although this collection is comprised of scans from several institutions, the ability of COVIDNet-CT to generalize to images from other countries, institutions, or CT imaging systems has not been evaluated. As such, COVIDNet-CT could be improved via additional training on a more diverse dataset.

\subsection{Architecture comparison}

    Next, we compare the performance of the proposed COVIDNet-CT with existing deep neural network architectures for the task of COVID-19 detection from chest CT images. More specifically, we compare it with the deep residual network architecture proposed in~\cite{resnetv2} (referred to here as ResNet-50), which is capable of achieving high accuracy, sensitivity, and PPV on the proposed COVIDx-CT benchmark dataset. It can be observed from Table~\ref{table:comp} that COVIDNet-CT achieves a test accuracy 0.4\% higher than that achieved with the ResNet-50 architecture while having 94.1\% fewer parameters and 90.2\% fewer FLOPs. Moreover, as shown in Table~\ref{table:sens} and Table~\ref{table:ppv} respectively, COVIDNet-CT achieves higher sensitivity and specificity than the ResNet-50 architecture across all infection types. These results highlight the benefits of leveraging machine-driven design exploration to create deep neural network architectures tailored to the task, data, and operational requirements. This is particularly relevant in clinical scenarios, as the ability to rapidly build and evaluate new deep neural network architectures is critical in order to adapt to changing data dynamics and operational requirements.

\subsection{Qualitative results}

\begin{figure}[ht]
        \centering
        \includegraphics[width=\textwidth]{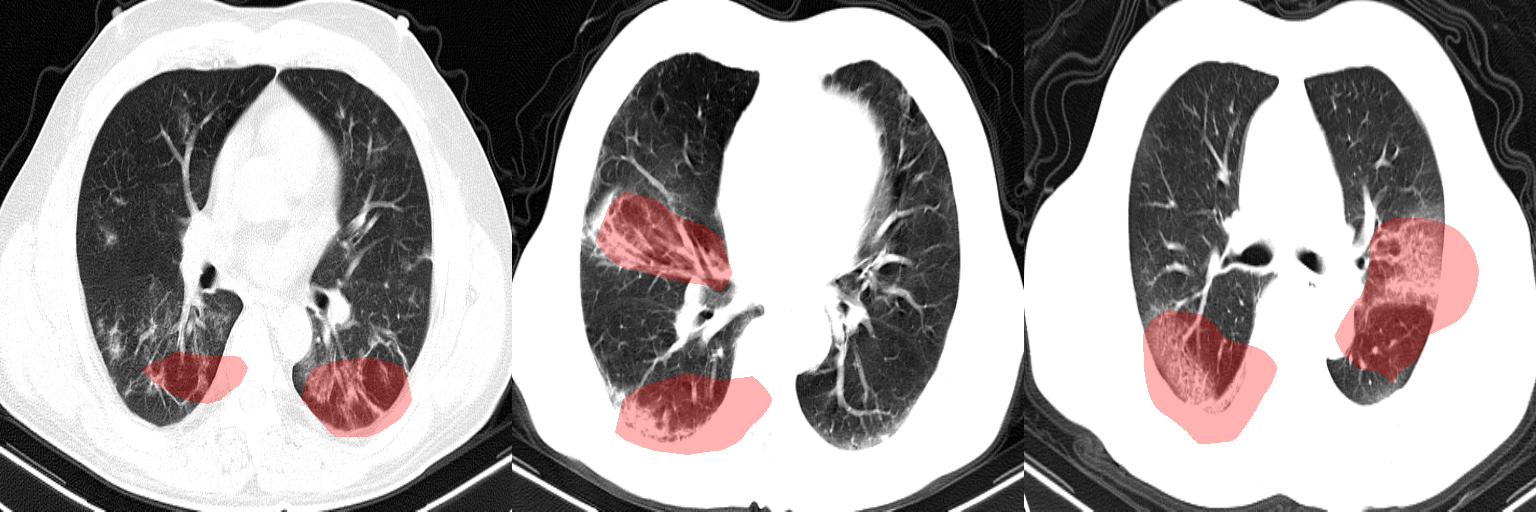}
        \caption{Example chest CT images of COVID-19 cases and their associated critical factors (highlighted in red) as identified by GSInquire~\cite{gsinquire}.}
        \label{fig:explain}
    \end{figure}

    In this study, we leveraged GSInquire~\cite{gsinquire} to perform explainability-driven performance validation of COVIDNet-CT in order to better understand its decision-making behaviour, and to ensure that its decisions are based on diagnostically-relevant imaging features rather than irrelevant visual indicators. Figure~\ref{fig:explain} shows the critical factors identified by GSInquire in three chest CT images of patients with COVID-19 pneumonia. Examining these visual interpretations, we observe that COVIDNet-CT primarily leverages abnormalities within the lungs in the chest CT images to identify COVID-19 cases, as well as to differentiate these cases from non-COVID-19 pneumonia cases. As previously mentioned, our initial experiments yielded deep neural networks that were found via explainability-driven performance validation to be basing their detection decisions on irrelevant indicators such as patient tables and imaging artifacts, which highlights the importance of leveraging explainability methods when building and evaluating deep neural networks for clinical applications. Furthermore, the ability to interpret how COVIDNet-CT detects COVID-19 cases may help clinicians trust its predictions, and may also help clinicians discover novel visual indicators of COVID-19 infection which could be leveraged in manual screening via CT imaging.

\section{Conclusion}

    In this study, we introduced COVIDNet-CT, a deep convolutional neural network architecture tailored for detection of COVID-19 cases from chest CT images via machine-driven design exploration. Additionally, we introduced COVIDx-CT, a benchmark CT image dataset consisting of 104,009 chest CT images across 1,489 patients. We quantitatively evaluated COVIDNet-CT using the COVIDx-CT test dataset in terms of accuracy, sensitivity, and PPV. Furthermore, we analysed the predictions of COVIDNet-CT via explainability-driven performance validation to ensure that its predictions are based on relevant image features and to better understand the CT image features associated with COVID-19 infection, which may aid clinicians in CT-based screening. In our analyses, we observed that COVIDNet-CT is highly performant when tested on the COVIDx-CT test dataset, and that abnormalities in the lungs are leveraged by COVIDNet-CT in its decision-making process.

    While COVIDNet-CT is not yet suitable for clinical use, we publicly released COVIDNet-CT and instructions for constructing the COVIDx-CT dataset as part of the COVID-Net open intiative in order to encourage broad usage and improvement by the research community. In the future, the performance and generalizability of COVIDNet-CT may be improved by expanding and diversifying the COVIDx-CT dataset, and COVIDNet-CT may also be extended to additional clinical tasks such as mortality risk stratification, lung function analysis, COVID-19 case triaging, and treatment planning. However, the ability to build solutions for these tasks is contingent on the availability of high-quality datasets. Finally, additional analysis of the explainability results may be performed in the future to identify key patterns in the CT images which may aid clinicians in manual screening.

\subsubsection*{Acknowledgments}

    We would like to thank Natural Sciences and Engineering Research Council of Canada (NSERC), the Canada Research Chairs program, CIFAR, DarwinAI Corp., NVIDIA Corp., and Hewlett Packard Enterprise Co.

\subsubsection*{Author contributions statement}

    H.G. and A.W. conceived the experiment, H.G., L.W., and A.W. conducted the experiment, H.G. and A.W. analysed the results. All authors reviewed the manuscript.

\subsubsection*{Additional information}

    \textbf{Competing interests:} L.W. and A.W. are affiliated with DarwinAI Corp.\\
    \textbf{Ethics Approval:} The study has received ethics clearance from the University of Waterloo (42235).

\medskip

\small

\bibliographystyle{IEEEtran}
\bibliography{references}

\begin{thebibliography}{10}
\providecommand{\url}[1]{#1}
\csname url@samestyle\endcsname
\providecommand{\newblock}{\relax}
\providecommand{\bibinfo}[2]{#2}
\providecommand{\BIBentrySTDinterwordspacing}{\spaceskip=0pt\relax}
\providecommand{\BIBentryALTinterwordstretchfactor}{4}
\providecommand{\BIBentryALTinterwordspacing}{\spaceskip=\fontdimen2\font plus
\BIBentryALTinterwordstretchfactor\fontdimen3\font minus
  \fontdimen4\font\relax}
\providecommand{\BIBforeignlanguage}[2]{{%
\expandafter\ifx\csname l@#1\endcsname\relax
\typeout{** WARNING: IEEEtran.bst: No hyphenation pattern has been}%
\typeout{** loaded for the language `#1'. Using the pattern for}%
\typeout{** the default language instead.}%
\else
\language=\csname l@#1\endcsname
\fi
#2}}
\providecommand{\BIBdecl}{\relax}
\BIBdecl

\bibitem{Wang2020_RTPCR}
W.~Wang, Y.~Xu, R.~Gao, R.~Lu, K.~Han, G.~Wu, and W.~Tan, ``Detection of
  sars-cov-2 in different types of clinical specimens,'' \emph{JAMA}, vol. 323,
  no.~18, pp. 1843--1844, 05 2020.

\bibitem{Yang2020}
Y.~Yang, M.~Yang, C.~Shen, F.~Wang, J.~Yuan, J.~Li, M.~Zhang, Z.~Wang, L.~Xing,
  J.~Wei, L.~Peng, G.~Wong, H.~Zheng, M.~Liao, K.~Feng, J.~Li, Q.~Yang,
  J.~Zhao, Z.~Zhang, L.~Liu, and Y.~Liu, ``Evaluating the accuracy of different
  respiratory specimens in the laboratory diagnosis and monitoring the viral
  shedding of 2019-ncov infections,'' \emph{medRxiv}, 2020.

\bibitem{Li2020_RTPCR}
Y.~Li, L.~Yao, J.~Li, L.~Chen, Y.~Song, Z.~Cai, and C.~Yang, ``Stability issues
  of rt-pcr testing of sars-cov-2 for hospitalized patients clinically
  diagnosed with covid-19,'' \emph{Journal of Medical Virology}, vol.~92,
  no.~7, pp. 903--908, 2020.

\bibitem{Ai2020}
T.~Ai, Z.~Yang, H.~Hou, C.~Zhan, C.~Chen, W.~Lv, Q.~Tao, Z.~Sun, and L.~Xia,
  ``Correlation of chest ct and rt-pcr testing for coronavirus disease 2019
  (covid-19) in china: A report of 1014 cases,'' \emph{Radiology}, vol. 296,
  no.~2, pp. E32--E40, 2020, pMID: 32101510.

\bibitem{Fang2020}
Y.~Fang, H.~Zhang, J.~Xie, M.~Lin, L.~Ying, P.~Pang, and W.~Ji, ``Sensitivity
  of chest ct for covid-19: Comparison to rt-pcr,'' \emph{Radiology}, vol. 296,
  no.~2, pp. E115--E117, 2020, pMID: 32073353.

\bibitem{Xie2020}
X.~Xie, Z.~Zhong, W.~Zhao, C.~Zheng, F.~Wang, and J.~Liu, ``Chest ct for
  typical coronavirus disease 2019 (covid-19) pneumonia: Relationship to
  negative rt-pcr testing,'' \emph{Radiology}, vol. 296, no.~2, pp. E41--E45,
  2020, pMID: 32049601.

\bibitem{ACR}
A.~C. of~Radiology, ``Acr recommendations for the use of chest radiography and
  computed tomography (ct) for suspected covid-19 infection,'' 2020.

\bibitem{Tian}
T.~et~al., ``Pulmonary pathology of early-phase 2019 novel coronavirus
  (covid-19) pneumonia in two patients with lung cancer,'' \emph{Journal of
  Thoracic Oncology}, 2020.

\bibitem{Shatri}
J.~Shatri, L.~Tafilaj, A.~Turkaj, K.~Dedushi1, M.~Shatri, S.~Bexheti, and S.~K.
  Mucaj, ``The role of chest computed tomography in asymptomatic patients of
  positive coronavirus disease 2019: A case and literature review,''
  \emph{Journal of Clinical Imaging Science}, 2020.

\bibitem{SinghE455}
\BIBentryALTinterwordspacing
N.~Singh and J.~Fratesi, ``Chest ct imaging of an early canadian case of
  covid-19 in a 28-year-old man,'' \emph{CMAJ}, vol. 192, no.~17, pp.
  E455--E455, 2020. [Online]. Available:
  \url{https://www.cmaj.ca/content/192/17/E455}
\BIBentrySTDinterwordspacing

\bibitem{Guan2020}
W.-j. Guan, Z.-y. Ni, Y.~Hu, W.-h. Liang, C.-q. Ou, J.-x. He, L.~Liu, H.~Shan,
  C.-l. Lei, D.~S. Hui, B.~Du, L.-j. Li, G.~Zeng, K.-Y. Yuen, R.-c. Chen, C.-l.
  Tang, T.~Wang, P.-y. Chen, J.~Xiang, S.-y. Li, J.-l. Wang, Z.-j. Liang, Y.-x.
  Peng, L.~Wei, Y.~Liu, Y.-h. Hu, P.~Peng, J.-m. Wang, J.-y. Liu, Z.~Chen,
  G.~Li, Z.-j. Zheng, S.-q. Qiu, J.~Luo, C.-j. Ye, S.-y. Zhu, and N.-s. Zhong,
  ``Clinical characteristics of coronavirus disease 2019 in china,'' \emph{New
  England Journal of Medicine}, vol. 382, no.~18, pp. 1708--1720, 2020.

\bibitem{Wang2020}
D.~Wang, B.~Hu, C.~Hu, F.~Zhu, X.~Liu, J.~Zhang, B.~Wang, H.~Xiang, Z.~Cheng,
  Y.~Xiong, Y.~Zhao, Y.~Li, X.~Wang, and Z.~Peng, ``Clinical characteristics of
  138 hospitalized patients with 2019 novel coronavirus–infected pneumonia in
  wuhan, china,'' \emph{JAMA}, vol. 323, no.~11, pp. 1061--1069, 03 2020.

\bibitem{Chung2020}
M.~Chung, A.~Bernheim, X.~Mei, N.~Zhang, M.~Huang, X.~Zeng, J.~Cui, W.~Xu,
  Y.~Yang, Z.~A. Fayad, A.~Jacobi, K.~Li, S.~Li, and H.~Shan, ``Ct imaging
  features of 2019 novel coronavirus (2019-ncov),'' \emph{Radiology}, vol. 295,
  no.~1, pp. 202--207, 2020, pMID: 32017661.

\bibitem{Pan2020}
F.~Pan, T.~Ye, P.~Sun, S.~Gui, B.~Liang, L.~Li, D.~Zheng, J.~Wang, R.~L.
  Hesketh, L.~Yang, and C.~Zheng, ``Time course of lung changes at chest ct
  during recovery from coronavirus disease 2019 (covid-19),'' \emph{Radiology},
  vol. 295, no.~3, pp. 715--721, 2020, pMID: 32053470.

\bibitem{Bai2020_Perf}
H.~X. Bai, B.~Hsieh, Z.~Xiong, K.~Halsey, J.~W. Choi, T.~M.~L. Tran, I.~Pan,
  L.-B. Shi, D.-C. Wang, J.~Mei, X.-L. Jiang, Q.-H. Zeng, T.~K. Egglin, P.-F.
  Hu, S.~Agarwal, F.-F. Xie, S.~Li, T.~Healey, M.~K. Atalay, and W.-H. Liao,
  ``Performance of radiologists in differentiating covid-19 from non-covid-19
  viral pneumonia at chest ct,'' \emph{Radiology}, vol. 296, no.~2, pp.
  E46--E54, 2020, pMID: 32155105.

\bibitem{Mei2020}
X.~Mei, H.-C. Lee, K.-y. Diao, M.~Huang, B.~Lin, C.~Liu, Z.~Xie, Y.~Ma,
  P.~Robson, M.~Chung, A.~Bernheim, V.~Mani, C.~Calcagno, K.~Li, S.~Li,
  H.~Shan, J.~Lv, T.~Zhao, J.~Xia, and Y.~Yang, ``Artificial
  intelligence–enabled rapid diagnosis of patients with covid-19,''
  \emph{Nature Medicine}, pp. 1--5, 05 2020.

\bibitem{cncb}
K.~Zhang, X.~Liu, J.~Shen, Z.~Li, Y.~Sang, X.~Wu, Y.~Zha, W.~Liang, C.~Wang,
  K.~Wang, L.~Ye, M.~Gao, Z.~Zhou, L.~Li, J.~Wang, Z.~Yang, H.~Cai, J.~Xu,
  L.~Yang, W.~Cai, W.~Xu, S.~Wu, W.~Zhang, S.~Jiang, L.~Zheng, X.~Zhang,
  L.~Wang, L.~Lu, J.~Li, H.~Yin, W.~Wang, O.~Li, C.~Zhang, L.~Liang, T.~Wu,
  R.~Deng, K.~Wei, Y.~Zhou, T.~Chen, J.~Y.-N. Lau, M.~Fok, J.~He, T.~Lin,
  W.~Li, and G.~Wang, ``Clinically applicable ai system for accurate diagnosis,
  quantitative measurements, and prognosis of covid-19 pneumonia using computed
  tomography,'' \emph{Cell}, vol.~18, no.~6, pp. 1423--1433, 2020.

\bibitem{covidnet}
L.~Wang and A.~Wong, ``Covid-net: A tailored deep convolutional neural network
  design for detection of covid-19 cases from chest x-ray images,'' 2020.

\bibitem{alex2020covidnets}
A.~Wong, Z.~Q. Lin, L.~Wang, A.~G. Chung, B.~Shen, A.~Abbasi,
  M.~Hoshmand-Kochi, and T.~Q. Duong, ``Covidnet-s: Towards computer-aided
  severity assessment via training and validation of deep neural networks for
  geographic extent and opacity extent scoring of chest x-rays for sars-cov-2
  lung disease severity,'' 2020.

\bibitem{Xu2020}
X.~Xu, X.~Jiang, C.~Ma, P.~Du, X.~Li, S.~Lv, L.~Yu, Q.~Ni, Y.~Chen, J.~Su,
  G.~Lang, Y.~Li, H.~Zhao, J.~Liu, K.~Xu, L.~Ruan, J.~Sheng, Y.~Qiu, W.~Wu,
  T.~Liang, and L.~Li, ``A deep learning system to screen novel coronavirus
  disease 2019 pneumonia,'' \emph{Engineering}, 2020.

\bibitem{Bai2020_Aug}
H.~X. Bai, R.~Wang, Z.~Xiong, B.~Hsieh, K.~Chang, K.~Halsey, T.~M.~L. Tran,
  J.~W. Choi, D.-C. Wang, L.-B. Shi, J.~Mei, X.-L. Jiang, I.~Pan, Q.-H. Zeng,
  P.-F. Hu, Y.-H. Li, F.-X. Fu, R.~Y. Huang, R.~Sebro, Q.-Z. Yu, M.~K. Atalay,
  and W.-H. Liao, ``Ai augmentation of radiologist performance in
  distinguishing covid-19 from pneumonia of other etiology on chest ct,''
  \emph{Radiology}, vol.~0, no.~0, p. 201491, 0, pMID: 32339081.

\bibitem{Li2020}
L.~Li, L.~Qin, Z.~Xu, Y.~Yin, X.~Wang, B.~Kong, J.~Bai, Y.~Lu, Z.~Fang,
  Q.~Song, K.~Cao, D.~Liu, G.~Wang, Q.~Xu, X.~Fang, S.~Zhang, J.~Xia, and
  J.~Xia, ``Using artificial intelligence to detect covid-19 and
  community-acquired pneumonia based on pulmonary ct: Evaluation of the
  diagnostic accuracy,'' \emph{Radiology}, vol. 296, no.~2, pp. E65--E71, 2020,
  pMID: 32191588.

\bibitem{Ardakani2020}
A.~A. Ardakani, A.~R. Kanafi, U.~R. Acharya, N.~Khadem, and A.~Mohammadi,
  ``Application of deep learning technique to manage covid-19 in routine
  clinical practice using ct images: Results of 10 convolutional neural
  networks,'' \emph{Computers in Biology and Medicine}, vol. 121, p. 103795,
  2020.

\bibitem{Shah2020}
V.~Shah, R.~Keniya, A.~Shridharani, M.~Punjabi, J.~Shah, and N.~Mehendale,
  ``Diagnosis of covid-19 using ct scan images and deep learning techniques,''
  \emph{medRxiv}, 2020.

\bibitem{Chen2020}
J.~Chen, L.~Wu, J.~Zhang, L.~Zhang, D.~Gong, Y.~Zhao, S.~Hu, Y.~Wang, X.~Hu,
  B.~Zheng, K.~Zhang, H.~Wu, Z.~Dong, Y.~Xu, Y.~Zhu, X.~Chen, L.~Yu, and H.~Yu,
  ``Deep learning-based model for detecting 2019 novel coronavirus pneumonia on
  high-resolution computed tomography: a prospective study,'' \emph{medRxiv},
  2020.

\bibitem{Zheng2020}
C.~Zheng, X.~Deng, Q.~Fu, Q.~Zhou, J.~Feng, H.~Ma, W.~Liu, and X.~Wang, ``Deep
  learning-based detection for covid-19 from chest ct using weak label,''
  \emph{medRxiv}, 2020.

\bibitem{Jin2020}
S.~Jin, B.~Wang, H.~Xu, C.~Luo, L.~Wei, W.~Zhao, X.~Hou, W.~Ma, Z.~Xu,
  Z.~Zheng, W.~Sun, L.~Lan, W.~Zhang, X.~Mu, C.~Shi, Z.~Wang, J.~Lee, Z.~Jin,
  M.~Lin, H.~Jin, L.~Zhang, J.~Guo, B.~Zhao, Z.~Ren, S.~Wang, Z.~You, J.~Dong,
  X.~Wang, J.~Wang, and W.~Xu, ``Ai-assisted ct imaging analysis for covid-19
  screening: Building and deploying a medical ai system in four weeks,''
  \emph{medRxiv}, 2020.

\bibitem{Jin2020_2}
C.~Jin, W.~Chen, Y.~Cao, Z.~Xu, Z.~Tan, X.~Zhang, L.~Deng, C.~Zheng, J.~Zhou,
  H.~Shi, and J.~Feng, ``Development and evaluation of an ai system for
  covid-19 diagnosis,'' \emph{medRxiv}, 2020.

\bibitem{Song2020}
Y.~Song, S.~Zheng, L.~Li, X.~Zhang, X.~Zhang, Z.~Huang, J.~Chen, H.~Zhao,
  Y.~Jie, R.~Wang, Y.~Chong, J.~Shen, Y.~Zha, and Y.~Yang, ``Deep learning
  enables accurate diagnosis of novel coronavirus (covid-19) with ct images,''
  \emph{medRxiv}, 2020.

\bibitem{Wang2020_Screen}
S.~Wang, B.~Kang, J.~Ma, X.~Zeng, M.~Xiao, J.~Guo, M.~Cai, J.~Yang, Y.~Li,
  X.~Meng, and B.~Xu, ``A deep learning algorithm using ct images to screen for
  corona virus disease (covid-19),'' \emph{medRxiv}, 2020.

\bibitem{gradcam}
R.~R. {Selvaraju}, M.~{Cogswell}, A.~{Das}, R.~{Vedantam}, D.~{Parikh}, and
  D.~{Batra}, ``Grad-cam: Visual explanations from deep networks via
  gradient-based localization,'' in \emph{2017 IEEE International Conference on
  Computer Vision (ICCV)}, 2017, pp. 618--626.

\bibitem{gensynth}
A.~Wong, M.~J. Shafiee, B.~Chwyl, and F.~Li, ``Ferminets: Learning generative
  machines to generate efficient neural networks via generative synthesis,''
  2018.

\bibitem{wong2019netscore}
\BIBentryALTinterwordspacing
A.~Wong, ``Netscore: Towards universal metrics for large-scale performance
  analysis of deep neural networks for practical usage,'' \emph{CoRR}, vol.
  abs/1806.05512, 2018. [Online]. Available:
  \url{http://arxiv.org/abs/1806.05512}
\BIBentrySTDinterwordspacing

\bibitem{resnet}
K.~He, X.~Zhang, S.~Ren, and J.~Sun, ``Deep residual learning for image
  recognition,'' in \emph{2016 IEEE Conference on Computer Vision and Pattern
  Recognition (CVPR)}, 2016, pp. 770--778.

\bibitem{resnetv2}
------, ``Identity mappings in deep residual networks,'' in \emph{Computer
  Vision - ECCV 2016}, B.~Leibe, J.~Matas, N.~Sebe, and M.~Welling, Eds.\hskip
  1em plus 0.5em minus 0.4em\relax Cham: Springer International Publishing,
  2016, pp. 630--645.

\bibitem{imagenet}
J.~{Deng}, W.~{Dong}, R.~{Socher}, L.~{Li}, {Kai Li}, and {Li Fei-Fei},
  ``Imagenet: A large-scale hierarchical image database,'' in \emph{2009 IEEE
  Conference on Computer Vision and Pattern Recognition}, 2009, pp. 248--255.

\bibitem{momentum}
N.~Qian, ``On the momentum term in gradient descent learning algorithms,''
  \emph{Neural Networks}, vol.~12, no.~1, pp. 145--151, 1999.

\bibitem{tensorflow}
M.~Abadi, A.~Agarwal, P.~Barham, E.~Brevdo, Z.~Chen, C.~Citro, G.~S. Corrado,
  A.~Davis, J.~Dean, M.~Devin, S.~Ghemawat, I.~Goodfellow, A.~Harp, G.~Irving,
  M.~Isard, Y.~Jia, R.~Jozefowicz, L.~Kaiser, M.~Kudlur, J.~Levenberg,
  D.~Man\'{e}, R.~Monga, S.~Moore, D.~Murray, C.~Olah, M.~Schuster, J.~Shlens,
  B.~Steiner, I.~Sutskever, K.~Talwar, P.~Tucker, V.~Vanhoucke, V.~Vasudevan,
  F.~Vi\'{e}gas, O.~Vinyals, P.~Warden, M.~Wattenberg, M.~Wicke, Y.~Yu, and
  X.~Zheng, ``{TensorFlow}: Large-scale machine learning on heterogeneous
  systems,'' 2015, software available from tensorflow.org.

\bibitem{gsinquire}
Z.~Q. Lin, M.~J. Shafiee, S.~Bochkarev, M.~S. Jules, X.~Y. Wang, and A.~Wong,
  ``Do explanations reflect decisions? a machine-centric strategy to quantify
  the performance of explainability algorithms,'' 2019.

\end{thebibliography}

\end{document}